\documentclass{aa}
\usepackage{graphics}
\def\th{\thinspace}
\def\ni{\noindent}

\begin{document}
  
  \thesaurus{}

  \title{Structural Properties of s-Cepheid Velocity Curves
           \thanks{Based on observations collected at the European
                      Southern Observatory (La Silla, Chile) and at the 
                      Observatoire de Haute-Provence (France)}
                      }
  \subtitle{Constraining the location of the $\omega_4 = 2\omega_1$ resonance}

   \author{F. Kienzle\inst{1} 
      \and P. Moskalik\inst{2}
      \and D. Bersier\inst{3}
      \and F. Pont\inst{1}}

   \institute{Observatoire de Geneve,
              Ch. des Maillettes 51,  
              CH-1290 Sauverny,
              Switzerland
             \and
	      Copernicus Astronomical Center, 
	      ul. Bartycka 18, 
	      00-716, Warsaw, 
	      Poland
	     \and
	      Mt. Stromlo and Siding Spring Observatories,
	      Weston Creek, 
	      ACT 2611, 
	      Australia}

   \offprints{F. Kienzle}
   \mail{Francesco.Kienzle@obs.unige.ch}

   \date{Received / Accepted}

   \maketitle
   \titlerunning{xx}
   \authorrunning{xx}

\begin{abstract}

The light curves of the first overtone Pop.\th I Cepheids (s-Cepheids) show
a discontinuity in their $\phi_{21}$ vs. $\mathrm{P}$ diagram, near
$\mathrm{P} = 3.2$\th day.  This feature, commonly attributed to the 2:1
resonance between the first and the fourth overtones ($\omega_4\approx
2\omega_1$), is not reproduced by the hydrodynamical models.  With the goal
of reexamining the resonance hypothesis, we have obtained new CORAVEL
radial velocity curves for 14 overtone Cepheids. Together with 10 objects
of Krzyt et~al. (\cite{krzyt}), the combined sample covers the whole range
of overtone Cepheid periods. The velocity Fourier parameters display a
strong characteristic resonant behavior. In striking contrast to
photometric ones, they vary smoothly with the pulsation period and show no
jump at 3.2\th day. The existing radiative hydrodynamical models match the
velocity parameters very well. The center of the $\omega_4 = 2\omega_1$
resonance is estimated to occur at $\mathrm{P}_\mathrm{r} = 4.58\pm
0.04$\th day, i.e.  at a period considerably longer than previously assumed
(3.2\th day). We identify two new members of the s-Cepheid group: \object{MY~Pup}
and \object{V440~Per}.

\end{abstract}

\section{Introduction} 

The {\it sinusoidal} or s-Cepheids constitute about 30\% of all Galactic
Cepheids with periods below 5\th days. Originally, they have been
discriminated from other Cepheids by qualitative criteria -- their small
amplitudes and almost sinusoidal light, color and radial velocity curves. A
more precise, quantitative definition, based on the Fourier decomposition
of the light curves, has been introduced by Antonello et~al.(\cite{antonello4}).
 Armed with this new classification tool, the Italian
group has identified 33 s-Cepheids, plus several likely suspects (Antonello
\& Poretti \cite{antonello1}; Antonello et~al. \cite{antonello4}; Mantegazza 
\& Poretti \cite{mantegazza}; Poretti \cite{poretti}).

The physical nature of the s-Cepheids has been a matter of debate. It has
been suggested in the General Catalog of Variable Stars that these stars
are either fundamental-mode pulsators during the first crossing of the
instability strip or, alternatively, first overtone pulsators. The latter
view has been adopted by Antonello et~al. (\cite{antonello4}).  A different
interpretation has been proposed by Gieren et~al.  (\cite{gieren}), who has
argued that the short period s-Cepheids pulsate in the first overtone, but
that the long period ones ($\mathrm{P} > 3.2$\th day) are in fact
fundamental-mode variables. The controversy has finally been settled with
the massive photometry of the MACHO and EROS, which has unambiguously shown
that {\it all} s-Cepheids pulsate in the overtone (Welch et~al.
\cite{welch}; Beaulieu et~al. \cite{beaulieu}). Independently, the first 
crossing hypothesis has been ruled out by the new spectroscopic abundance
analysis (Kovtyukh et~al. \cite{kovtyukh}).

The debate between Antonello et~al. and Gieren et~al. has been sparked off
by the curious behavior of the s-Cepheid Fourier parameters. The light
curve Fourier phase $\phi_{21}$ (cf. Simon \& Lee \cite{simon1}), when
plotted vs. pulsation period, shows a very deep and very abrupt drop in the
vicinity of $\mathrm{P}=3.2$\th day.  At the same period the amplitude
ratios $\mathrm{R}_{21}$ and $\mathrm{R}_{31}$ display a pronounced
minimum. This behavior is reminiscent of what is observed for the
fundamental-mode Cepheids at $\mathrm{P}\approx 10$\th day (Simon \&
Moffett \cite{simon2}).  In the latter case the characteristic variation of
the Fourier parameters has its origin in the 2:1 resonance between the
fundamental mode and the second overtone (Simon \& Schmidt \cite{simon};
Buchler et~al. \cite{BMK}; Kov\'acs \& Buchler \cite{kovacs}).

By analogy, Antonello \& Poretti (\cite{antonello1}) and Petersen
(\cite{petersen2}) have proposed that the variations observed for the
s-Cepheid light curves are also caused by a resonance, namely the 2:1
coupling between the first and the fourth overtones.  Also by analogy, it
has been assumed that the resonance center coincides with the drop of the
photometric $\phi_{21}$ and therefore occurs at $\mathrm{P}=3.2$\th day.

The resonance hypothesis, although very attractive, has encountered serious
difficulties, when confronted with hydrodynamical calculations.  Three sets
of overtone cepheid models have been specifically computed to study the
presumed resonance (Antonello \& Aikawa \cite{antonello2}, \cite{antonello3}; 
Schaller \& Buchler \cite{schaller}). To great disappointment, they have
all failed to reproduce the properties of the s-Cepheid light curves. The
theoretical $\phi_{21}$ and $\mathrm{R}_{21}$ display some features in the
vicinity of $\mathrm{P}=3.2$\th day but they are {\it very far} from
reproducing what is actually observed.  The discrepancy is even more
embarrassing, when compared with the very good agreement obtained with the
same codes for the fundamental-mode pulsators (e.g. Moskalik et~al. \cite{MBM}).

Another potential problem has been pointed out by Buchler et~al.
(\cite{buchler}), who have considered the constraints imposed by resonances
on the evolutionary Mass-Luminosity relation. Their linear calculations
show that the proposed s-Cepheid resonance centered at 3.2\th day and the
well established f-mode Cepheid resonance at 10\th day cannot be reconciled
simultaneously with the same $\mathrm{M}-\mathrm{L}$ relation.  For
consistent picture, the s-Cepheid resonance has to occur at
$\mathrm{P}=4.3$\th day.

So far, the analysis of the s-Cepheid pulsations has been performed almost
exclusively in the photometry domain. This choice has been dictated by the
lack of high quality $V_r$ data for these low amplitude stars.  The only
attempt to compare the s-Cepheid velocity curves with the hydrodynamical
models has been largely inconclusive (Antonello \& Aikawa \cite{antonello3}).
Analyzing overtone Cepheid velocity data is particularly desirable in light
of the modeling difficulties discussed above. Velocity Fourier parameters
can provide additional information on the overtone Cepheid pulsation
dynamics and, thus, can shed new light on the resonance puzzle.  The radial
velocity, being a dynamical quantity, should display the effects of 
resonances in a more visible way. The models of BL~Her-type stars give a
good example of such a behavior (Moskalik \& Buchler \cite{moskalik2},
Buchler \& Buchler \cite{buchler3}). The use of the velocity data is also
preferred for comparison with the hydrodynamical computations, which are
known to reproduce the observed velocity curves very well (e.g. Moskalik 
et~al. \cite{MBM}). As for the light curves, the models show
small but persistent discrepancies of the Fourier phases, $\phi_{n1}$, for
every type of radial pulsators studied (Simon \& Aikawa \cite{simon3},
Simon \cite{simon4}, Moskalik et~al. \cite{MBM}, Moskalik \&
Buchler \cite{moskalik2}). The problem is most likely caused by an
inadequate treatment of the radiative transfer in outer stellar layers,
which however, has very little effect on the computed velocity curves
(Feuchtinger \& Dorfi \cite{feuchtinger}).

With the above reasoning in mind, we have collected new CORAVEL radial
velocity data for several known overtone cepheids. Several others have
recently been analyzed by Krzyt et~al. (\cite{krzyt}).  The combined sample
for the first time gives a complete and accurate description of the entire
s-Cepheid velocity Fourier progression. The preliminary results of this
project have been presented by Kienzle et~al. (\cite{kienzle}).  In this
paper, we present the final results and discuss their astrophysical
implications. In particular, we discuss the constraint imposed by the
velocity data on the location of the s-Cepheid resonance.

\section{Observations}

Krzyt et~al. (\cite{krzyt}) have made an extensive compilation of published
radial velocity measurements for classical Cepheids and subsequently used
these data to derive accurate Fourier parameters of their pulsation
velocity curves. Their sample, over 100 objects in total, contains,
however, only 10 overtone Cepheids. This number is not sufficient to
adequately cover the Fourier progression for this group of stars.
Therefore, we have taken new data for 13 other known overtone Cepheids, in
order to enlarge the sample of Krzyt et~al. The number of selected targets
is particularly large close to the photometric $\phi_{21}$ drop (i.e. close
to $\mathrm{P} = 3.2$\th day), where the resonance has been expected,
according to the previous results.  In addition, 3 other Cepheids have been
observed (\object{AP~Pup}, \object{MY~Pup} and \object{\object{IT~Car}}),
in an attempt to identify new long period overtone pulsators.  Overtone
Cepheids with $\mathrm{P}>5$\th day have recently been found in the LMC
(Alcock et~al. \cite{alcock}, Fig.~5), their existence in the Galaxy has
also been predicted on theoretical grounds (Buchler et~al. \cite{buchler2}).

The $V_{r}$ observations have been obtained with the northern and southern
CORAVEL cross-correlation spectrophotometer (Baranne et~al. \cite{baranne})
 at the 1-m Swiss telescope at the Haute Provence Observatory 
(France) and at the 1.54-m Danish telescope at the European Southern
Observatory, La Silla (Chile). The cross-correlation function has been
fitted with a gaussian profile in a standard way (Burki et~al.
\cite{burki}) in order to extract the radial velocities. Radial velocity 
standards have also been observed to check the instrumental drift. The
majority of the data has been collected during four runs; in December 1996
(by FP), February 1997 (by FK), June 1997 (by DB) and during the last
southern CORAVEL run at La Silla, in December 1997 (by FK).  For all
program Cepheids a very good phase coverage has been achieved, with more
than 30 points per star (except \object{V379~Cas} -- 27 points). The
measurement errors range from 0.3~km/s to 0.8~km/s in most cases.

The data and Table 1 hereafter are available at the CDS \footnote{anonymous
ftp to cdsarc.u.strasbg.fr (130.79.128.5);
http://cdsweb.u-strasbg.fr/Abstract.html}.\\

\section{Fourier decomposition}

The radial velocity data are fitted with 

$$V_{r}(t)=\mathrm{A}_0+\sum_{k=1}^{N} \mathrm{A}_k \sin[k\omega(t-t_0) +
\phi_k
]\eqno(1)$$ 

\ni where $\omega=2\pi/\mathrm{P}$, $\mathrm{P}$ is the pulsation period of 
the star and $N$ is the order of the fit.  The parameters $\mathrm{A}_0$,
$\mathrm{A}_k$, $\phi_k$, $\mathrm{P}$ and their errors are estimated with
a standard unweighted least-squares method.  The variance of the residuals,
$\sigma^{2}$, for $M$ data points is estimated as:

$$\sigma^2 = \frac{\chi^{2}_{\mathrm{min}}}{M-2N-2}\eqno(2)$$

\begin{figure}
  \resizebox{\hsize}{!}{\includegraphics{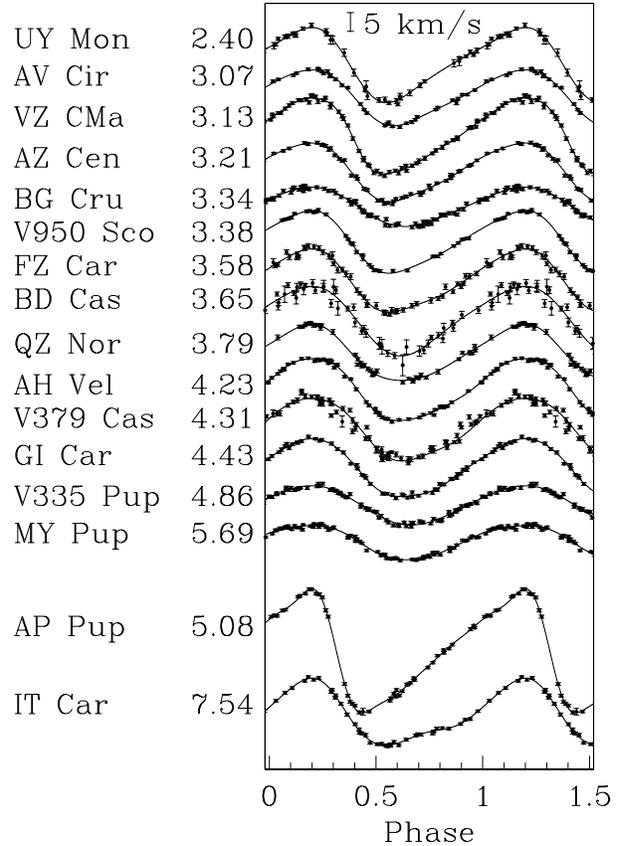}}
  \caption{Radial velocity curves for our program Cepheids. Overtone pulsators
           are sorted by increasing period, from top to bottom. 
           Two fundamental-mode Cepheids, \object{AP~Pup} and \object{IT~Car}
           are shown at the bottom of the plot.}
  \label{fig:vr_curves}
\end{figure}

\ni where $\chi^2_{\rm min}$ is the sum of squared residuals.
The order of the fit, $N$, is increased until adding another 
harmonic does not decrease $\sigma$ significantly. The points which
are more than 2.5$\sigma$ away from the fit (these are assumed to be poor
quality data) are eliminated, and the fitting procedure is repeated.
As the last step, the Fourier phases $\phi_{k1}\equiv\phi_{k}-k\phi_{1}$
and amplitude ratios $\mathrm{R}_{k1}\equiv\mathrm{A}_{k}/\mathrm{A}_{1}$
are calculated.  Their errors are computed with the formulae of Petersen
(\cite{petersen1}).  In Fig.\th \ref{fig:vr_curves} we display the phased
velocity curves for the program stars, together with their Fourier fits.
The Fourier parameters and their formal errors are given in Table~1.

\begin{table*}
\begin{tabular}{|l|r|r|r|c|r|r|r|r|r|r|c|c|c|c|} \hline
Star&Period
[d]&$M$&$N$&$\sigma$&$A_{0}$&$A_{1}$&$R_{21}$&$\phi_{21}$&$R_{31}$&$\phi_{31}$&$R_{41}$&$\phi_{41}$&$R_{51}$&$\phi_{51}$\\
\hline
\hline
\object{UY Mon}   &2.398240 & 36& 3& 0.611& 33.750 &  9.136 &  0.276 &  2.968 &
 0.043 &  6.088 &--&--&--&--\\ 
         &0.000005 &   &  &      &  0.116 &  0.147 &  0.020 &  0.069 &  0.017 &
 0.415 &--&--&--&--\\ 
\object{AV Cir}   &3.065006 & 35& 3& 0.435&  4.787 &  7.128 &  0.203 &  3.116 &
 0.037 &  5.525 &--&--&--&--\\ 
         &0.000158 &   &  &      &  0.079 &  0.110 &  0.017 &  0.086 &  0.016 &
 0.401 &--&--&--&--\\ 
\object{VZ CMa}   &3.126326 & 56& 5& 0.488& 39.421 &  9.692 &  0.256 &  3.194 &
 0.080 &  6.079 & 0.038 & 2.821 & 0.021 & 5.484\\ 
         &0.000139 &   &  &      &  0.067 &  0.090 &  0.011 &  0.045 &  0.010 &
 0.130 & 0.010 & 0.272 & 0.009 & 0.480\\ 
\object{AZ Cen}   &3.211752 & 46& 3& 0.462&-10.343 &  7.585 &  0.198 &  3.153 &
 0.061 &  5.967 & --    & --    & --    & --   \\ 
         &0.000126 &   &  &      &  0.073 &  0.111 &  0.015 &  0.072 &  0.014 &
 0.207 & --    & --    & --    & --   \\ 
\object{BG Cru}   &3.342503 & 67& 2& 0.465&-19.708 &  5.178 &  0.108 &  3.171 &
 --    &  --    & --    & --    & --    & --   \\ 
         &0.000056 &   &  &      &  0.057 &  0.084 &  0.017 &  0.152 &  --    &
 --    & --    & --    & --    & --   \\ 
\object{V950 Sco} &3.382519 & 34& 3& 0.340& 13.687 &  7.962 &  0.214 &  3.266 &
 0.051 &  6.352 & --    & --    & --    & --   \\ 
         &0.000012 &   &  &      &  0.071 &  0.101 &  0.012 &  0.057 &  0.013 &
 0.216 & --    & --    & --    & --   \\ 
\object{FZ Car}   &3.577946 & 47& 3& 0.799& -0.960 &  8.110 &  0.211 &  3.492 &
 0.056 &  6.417 & --    & --    & --    & --   \\ 
         &0.000023 &   &  &      &  0.121 &  0.175 &  0.024 &  0.101 &  0.020 &
 0.394 & --    & --    & --    & --   \\ 
\object{BD Cas}   &3.650765 & 43& 2& 1.331&-49.084 &  8.996 &  0.153 &  2.834 &
 --    &  --    & --    & --    & --    & --   \\ 
         &0.000063 &   &  &      &  0.221 &  0.306 &  0.033 &  0.268 &  --    &
 --    & --    & --    & --    & --   \\ 
\object{QZ Nor}   &3.786553 & 43& 3& 0.536&-39.549 &  7.383 &  0.158 &  3.695 &
 0.044 &  7.615 & --    & --    & --    & --   \\ 
         &0.000031 &   &  &      &  0.084 &  0.147 &  0.018 &  0.150 &  0.022 &
 0.435 & --    & --    & --    & --   \\ 
\object{AH Vel}   &4.226777 & 50& 4& 0.410& 25.801 &  8.368 &  0.157 &  3.590 &
 0.042 &  5.873 & 0.023 & 3.304 & --    & --   \\ 
         &0.000053 &   &  &      &  0.070 &  0.085 &  0.009 &  0.086 &  0.011 &
 0.233 & 0.010 & 0.468 & --    & --   \\ 
\object{V379 Cas} &4.305816 & 27& 3& 0.669&-38.652 &  8.428 &  0.111 &  3.427 &
 0.055 &  5.533 & --    & --    & --    & --   \\ 
         &0.000453 &   &  &      &  0.131 &  0.186 &  0.024 &  0.197 &  0.023 &
 0.438 & --    & --    & --    & --   \\ 
\object{GI Car}   &4.431035 & 50& 4& 0.448&-20.204 &  7.819 &  0.119 &  3.458 &
 0.037 &  6.644 & 0.019 & 2.222 & --    & --   \\ 
         &0.000287 &   &  &      &  0.066 &  0.093 &  0.013 &  0.095 &  0.012 &
 0.318 &  0.012 & 0.644 & --    & --   \\ 
\object{V335 Pup} &4.860555 & 70& 2& 0.466& 40.041 &  5.249 &  0.062 &  3.036 &
 --    &  --    & --    & --    & --    & --   \\ 
         &0.000354 &   &  &      &  0.056 &  0.082 &  0.015 &  0.254 &  --    &
 --    & --    & --    & --    & --   \\ 
\object{MY Pup}   &5.694670 & 72& 3& 0.429& 15.168 &  4.714 &  0.080 &  2.641 &
 0.043 &  6.108 & --    & --    & --    & --   \\ 
         &0.000317 &   &  &      &  0.052 &  0.070 &  0.016 &  0.195 &  0.018 &
 0.335 & --    & --    & --    & --   \\ 
\hline                                                                        
\hline						                    
      
\object{AP Pup}   &5.084534 & 45& 5& 0.521& 15.393 & 14.016 &  0.369 &  3.115 &
 0.183 &  6.384 & 0.090 & 3.325 & 0.030 & 0.327\\
         &0.000090 &   &  &      &  0.098 &  0.125 &  0.009 &  0.031 &  0.009 &
 0.055 & 0.010 & 0.089 & 0.008 & 0.360\\
\object{IT Car}   &7.539680 & 40& 3& 0.402&-11.801 &  8.004 &  0.343 &  3.871 &
 0.033 &  7.016 & --    & --    & --    & --   \\
         &0.002797 &   &  &      &  0.068 &  0.098 &  0.013 &  0.042 &  0.011 &
 0.366 & --    & --    & --    & --   \\
\hline		    
\end{tabular}
\caption {Fourier parameters for Cepheids measured with CORAVEL. $M$, $N$ 
and $\sigma$ are, respectively, the number of datapoints, the order of the 
fit and the standard deviation of residuals.}
\label{tab:low_fourier}
\end{table*}

\subsection{Comments on individual stars}

For 4 stars the phase coverage has been improved by supplementing our CORAVEL
measurements with the published data:\\
{\it BD~Cas} -- 18 points from Gorynya et~al. (\cite{gorynya1}, 
                \cite{gorynya2})\\
{\it FZ~Car} -- 5 points from Pont et~al. (\cite{pont})\\
{\it UY~Mon} -- 18 points from Imbert (\cite{imbert}). The archival data have
                been re-reduced. A vertical shift of $-1.1$~km/s has been 
                applied, which reduces the variance of the fit, $\sigma^2$, 
                by 48\%.\\
{\it QZ~Nor} -- 5 points from Metzger et~al. (\cite{metzger}).
                A vertical shift of $-0.9$~km/s has been applied, which 
                reduced the variance of the fit by 25\%.\\

\section{Properties of Fourier parameters}

\begin{figure}
  \resizebox{\hsize}{!}{\includegraphics{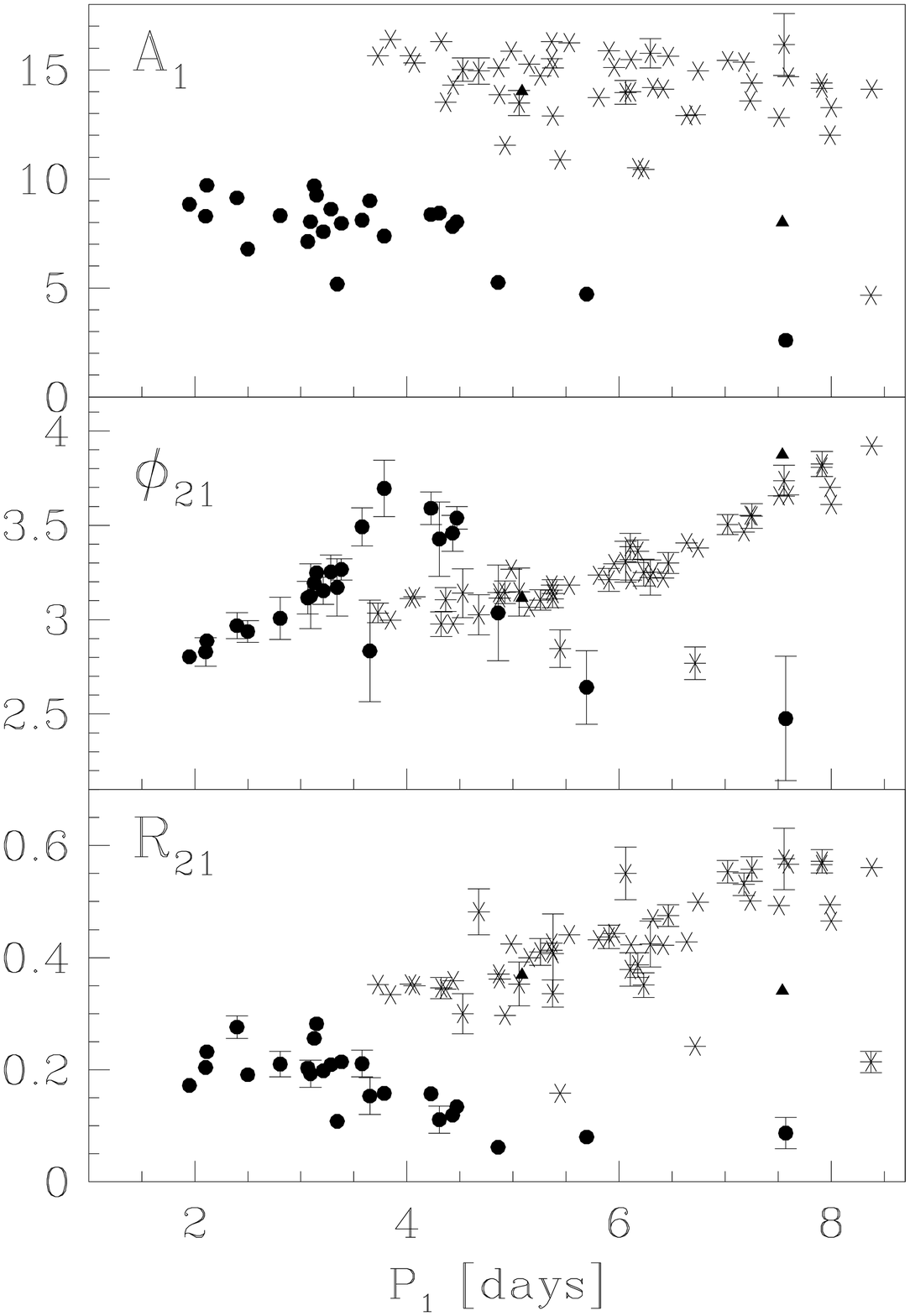}}  
  \caption{$\mathrm{A}_{1}$ (in km/s), $\phi_{21}$ and $\mathrm{R}_{21}$ vs. 
           period for Cepheid radial velocity curves (top, middle and bottom,
           respectively). Fundamental-mode Cepheids are marked with asterisks 
           and overtone Cepheids with filled circles. Filled triangles 
           represent \object{AP~Pup} and \object{IT~Car} (see text for these 
           two stars). Error bars are shown only when larger than the symbol 
           size.}
  \label{fig:A1_phi21_R21}
\end{figure}

\begin{figure}
  \resizebox{\hsize}{!}{\includegraphics{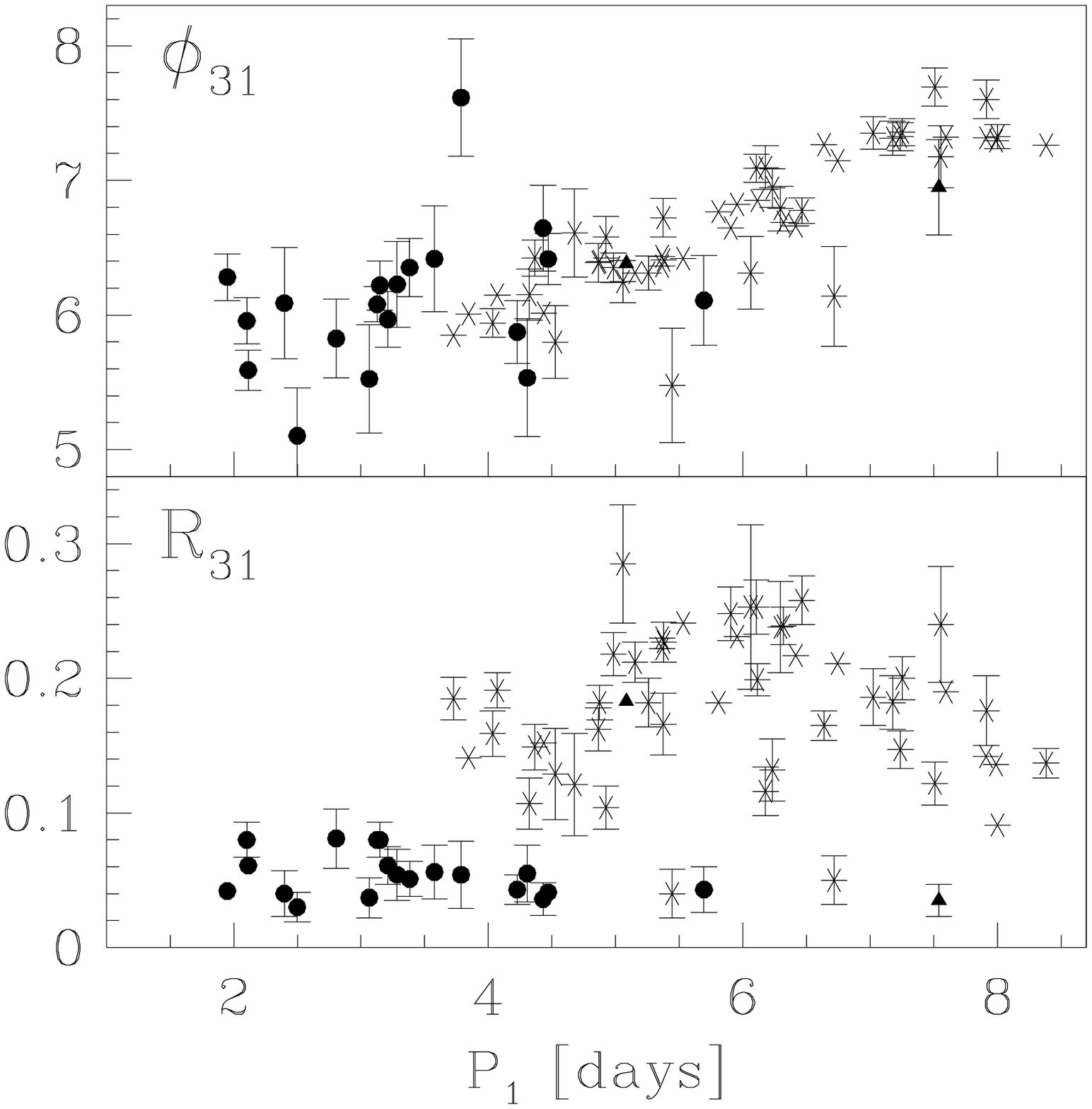}}
  \caption{$\phi_{31}$ and $\mathrm{R}_{31}$ vs. period for Cepheid radial 
           velocity curves (top and bottom, respectively). Same symbols as in 
           Fig.\th \ref{fig:A1_phi21_R21}.}
  \label{fig:phi31_R31}
\end{figure}

The set of s-Cepheids observed for this paper has been supplemented with 10
overtone pulsators analyzed by Krzyt et~al. (\cite{krzyt}): \object{SU~Cas},
\object{EU~Tau}, \object{IR~Cep}, \object{DT~Cyg}, \object{V351~Cep}, 
\object{EV~Sct}, \object{SZ~Tau}, \object{V532~Cyg}, \object{FF~Aql} and
\object{V440~Per}. The lowest order Fourier parameters $\mathrm{A}_{1}$,
$\phi_{21}$ and $\mathrm{R}_{21}$ for the entire sample are plotted vs. 
period in Fig.\th \ref{fig:A1_phi21_R21} (filled circles).  In the same
plot we also display Krzyt's et~al.  fundamental-mode Cepheids (asterisks).
In this case we have limited the sample to stars with the most secure
Fourier solutions, namely those with 25 or more datapoints and with an
error of $\sigma(\phi_{21}) < 0.15$. The third order parameters $\phi_{31}$
and $\mathrm{R}_{31}$ are displayed for completeness in Fig.\th
\ref{fig:phi31_R31}.


\smallskip

Before we begin the general discussion of Figs.\th \ref{fig:A1_phi21_R21} and 
\ref{fig:phi31_R31} preliminary clarifications for several stars are in order:

{\it \object{AP~Pup} and \object{IT~Car}} -- These stars are displayed in
the plots as filled triangles.  All the Fourier parameters place \object{AP~Pup} 
securely among the fundamental pulsators. The high values of
$\mathrm{A}_1$, $\mathrm{R}_{21}$ and $\mathrm{R}_{31}$ are incompatible
with it being an overtone Cepheids.  For \object{\object{IT~Car}},
$\mathrm{A}_1$ and $\mathrm{R}_{21}$ are somewhat low (and
$\mathrm{R}_{31}$ very low), but $\phi_{21}$ and $\phi_{31}$ fall on the
fundamental mode sequence. We classify both stars as fundamental-mode
Cepheids and will not discuss them any further here.

{\it \object{MY~Pup}} ($\mathrm{P}=5.69$\th day) -- The values of
$\mathrm{A}_1$, $\mathrm{R}_{21}$, $\mathrm{R}_{31}$ and $\phi_{21}$ for
this object are well below those derived for the fundamental-mode Cepheids.
On these grounds, we classify this star as a new overtone Cepheid.  This
conclusion will be strengthened by comparison with the hydrodynamical
models (Section~5). The light curve of \object{MY~Pup} has been Fourier
analyzed by Antonello \& Poretti (\cite{antonello1}).  Antonello et~al.
(\cite{antonello4}) have described it as a suspected s-Cepheid.

{\it \object{V440~Per}} ($\mathrm{P}=7.57$\th day) -- This very low
amplitude variable ($\mathrm{A}_1=2.7\pm 0.1$~km/s) has been analyzed by
Krzyt et~al.  (\cite{krzyt}). It is similar to \object{MY~Pup} and, for the
same reasons, we classify this star as a new overtone Cepheid. Its light
curve has been Fourier analyzed by Antonello \& Poretti
(\cite{antonello1}). The classification scheme of Antonello et~al.
(\cite{antonello4}), based on photometric $\phi_{21}$, does not
discriminate between the pulsation modes for $\mathrm{P} > 5.5$\th day (see
also Welch et~al. \cite{welch}, Fig.\th 2).  Therefore, the overtone nature
of V440~Per has not been recognized in their paper.

{\it \object{BD~Cas}} -- This star has originally been classified as
Pop.\th II Cepheid, according to its photometric and spectroscopic
characteristics (Petit \cite{petit}), although its low galactic latitude
($b=-0.96$) is also compatible with Pop.~I. With new CCD photometry,
Schmidt (\cite{schmidt}) and Poretti (\cite{poretti}) have reclassified
\object{BD~Cas} as a Pop.\th I s-Cepheid. However, their photometric data are
scarce (16 points) and the Fourier parameters are plagued with very large
errors ($\sigma(\phi_{2})=0.72$).  The values of $\mathrm{A}_1$ and
$\mathrm{R}_{21}$ derived from the velocity curve place the star among the
overtone Cepheids, but $\phi_{21} = 2.834\pm 0.268$ does not, being instead
similar to those of the fundamental pulsators. Clearly, more observations
are needed to confirm the s-Cepheid status of this variable.

{\it \object{X~Lac}, \object{V495~Cyg} and \object{V636~Cas}} -- these 3
variables have been analyzed by Krzyt et~al. (\cite{krzyt}). For 
\object{X~Lac} ($\mathrm{P}=5.45$\th day) and \object{V495~Cyg} 
($\mathrm{P}=6.72$\th day) the values of $\phi_{21}$, $\mathrm{R}_{21}$ and
$\mathrm{R}_{31}$ fall well below the fundamental mode sequence and close
to those of \object{MY~Pup} and \object{V440~Per}. This behavior strongly
suggests the s-Cepheid classification, but the amplitude of both stars is
rather high and typical of fundamental pulsators ($10.88\pm 0.16$~km/s and
$12.95\pm 0.24$~km/s, respectively). In the case of \object{V636~Cas}
($\mathrm{P}=8.38$\th day), the values of $\mathrm{A}_1 = 4.66\pm
0.08$~km/s and $\mathrm{R}_{21} = 0.214\pm 0.019$ point towards the
s-Cepheid classification, but the very high value of $\phi_{21} = 4.554\pm
0.087$ is in conflict with such an interpretation.  Although all 3 stars
differ from the majority of the fundamental-mode Cepheids, at this point
the evidence is not sufficient to consider them overtone pulsators.

\medskip

Our sample of 24 s-Cepheids covers the range of periods from 1.95 to
7.57\th day. Except for \object{BD~Cas} which deviates from the trend, the
remaining stars display a remarkably tight progression of $\phi_{21}$,
$\mathrm{R}_{21}$ and to a lesser degree of $\mathrm{A}_1$ with the
pulsation period. As the period increases, $\phi_{21}$ rises, reaches a
maximum at about $\mathrm{P}=4$\th day and then falls down to $\sim 2.5$\th
rad.  This variation is accompanied by a decrease of $\mathrm{R}_{21}$ at
periods of $3.5-5.0$\th day, followed by a very slow increase. The
amplitude $\mathrm{A}_1$ decreases uniformly throughout the whole range of
periods. In contrast to the low order parameters, the behavior of the
higher order terms $\phi_{31}$ and $\mathrm{R}_{31}$ shows essentially no
features, perhaps because of their low accuracy.

The velocity Fourier parameters of the overtone Cepheids are distinctively
different from those of the fundamental-mode pulsators. In particular,
$\mathrm{A}_1$, $\mathrm{R}_{21}$ and $\mathrm{R}_{31}$ are much lower,
which is a testimony to the ``sinusoidal'' shape of the s-Cepheid velocity
curves.  The two groups are also clearly separated in the $\phi_{21} -
\mathrm{P}$ plane, except a narrow range of periods around 5\th day. These 
properties allow to distinguish overtone from fundamental-mode Cepheids in
the {\it entire} range of periods, including $\mathrm{P} > 5.5$\th day,
where Antonello's et~al. (\cite{antonello4}) criterion based on the
light curve $\phi_{21}$ no longer works.

\begin{figure}
  \resizebox{\hsize}{!}{\includegraphics{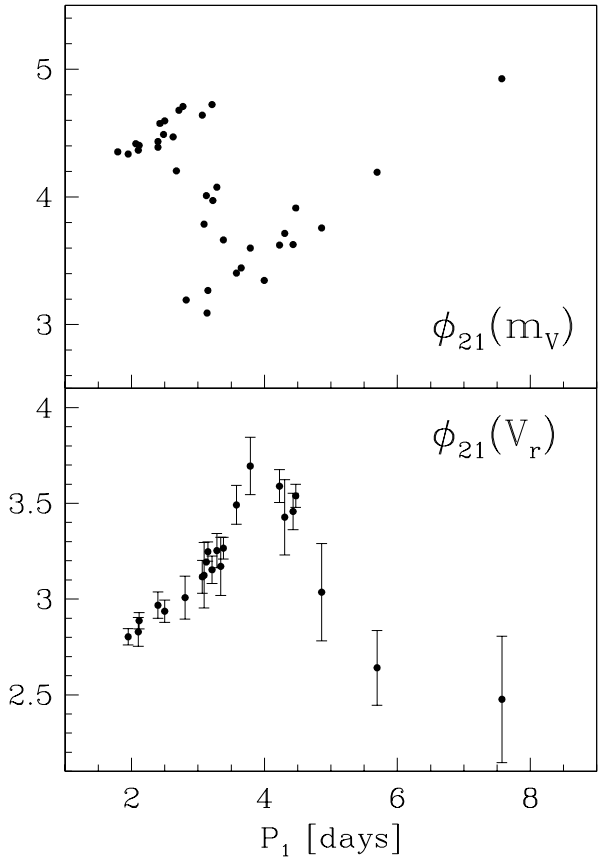}}
  \caption{s-Cepheid $\phi_{21}$ vs. period for light curves (top) and for 
           radial velocity curves (bottom). For 2 stars photometric 
           $\phi_{21}$ is not available: \object{BG~Cru} (first harmonic 
           not detectable) and V351~Cep (no published light curve Fourier
           decomposition). \object{BD~Cas} is omitted in both plots.}
  \label{fig:phi21}
\end{figure}

In Fig.\th \ref{fig:phi21} we compare the s-Cepheid $\phi_{21}$ progression
for light curves and for velocity curves. The light curve data are taken
from Antonello \& Poretti (\cite{antonello1}), Antonello~et.~al.
(\cite{antonello4}), Mantegazza \& Poretti (\cite{mantegazza}) and Poretti
(\cite{poretti}). The two $\phi_{21}$ plots are remarkably different. For
the light curves, $\phi_{21}$ undergoes a dramatic, essentially
discontinuous drop at a period of $\sim 3.2$\th day. It is this behavior,
accompanied by a local minimum of $\mathrm{R}_{21}$, which led to the
hypothesis that the 2:1 resonance between the first overtone and the fourth
overtone occurs at this place (Antonello \& Poretti \cite{antonello1};
Petersen \cite{petersen2}).  In case of the velocity curves, $\phi_{21}$
varies smoothly and displays {\it no jump at 3.2\th day} nor at any other
period.  The same is true for all the remaining velocity Fourier
parameters. In other words, there is no spectacular change of velocity
curve morphology at 3.2\th day.  This surprising and unexpected result
contradicts the assumption of a resonance occurring at this particular
period. The resonance does not have to cause discontinuous variations of
the Fourier parameters.  When it does, however, it happens for {\it both}
the velocity curves and the light curves (e.g. Buchler \& Kov\'acs
\cite{buchler1}). Because for the s-Cepheids this is not the case, we must 
conclude that the 3.2\th day feature in their light curves is not related
to the resonance.  We will show in the next section that the resonance is
nevertheless present in these stars, but its center is located at a very
different period.

\section{Comparison with hydrodynamical models}

Two extended surveys of nonlinear first overtone Cepheid models have been
performed in recent years, by Schaller \& Buchler (\cite{schaller}) and by
Antonello \& Aikawa (\cite{antonello3}). Those surveys have been aimed at
investigating the effects of the $\omega_4\approx 2\omega_1$ resonance.  In
order to study these effects in a systematic fashion, the models have been
grouped in one parameter sequences, running either at constant luminosity
(Antonello \& Aikawa) or parallel to the theoretical first overtone Blue
Edge (Schaller \& Buchler).  Both sets of calculations are performed with
purely radiative hydrocodes and almost the same input physics (e.g. opacity
tables).  Consequently, they both give very similar results. In the
following discussion we will use models of Schaller \& Buchler,
primarily because their sequences cover a wider range of period ratios,
$\mathrm{P}_4/\mathrm{P}_1$.

\begin{figure}
  \resizebox{\hsize}{!}{\includegraphics{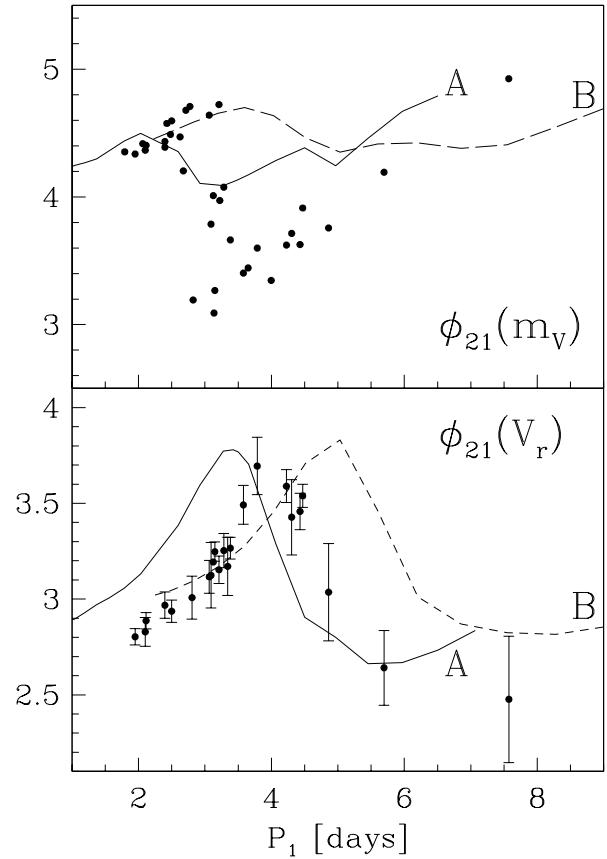}}
  \caption{Same as Fig.\th \ref{fig:phi21}, but compared with two sequences 
           of hydrodynamical models of Schaller \& Buchler (\cite{schaller}).
           See text for details.}
  \label{fig:phi21a}
\end{figure}

In Fig.\th \ref{fig:phi21a} we plot the observational data together with
the theoretical values of $\phi_{21}$ for two sequences of models. Sequence
A (solid line) obeys the Mass-Luminosity relation of Chiosi (\cite{chiosi})
and runs on the H--R diagram parallel to, but 100K cooler of the first
overtone Blue Edge.  Sequence B (dashed line) follows the classical
$\mathrm{M}-\mathrm{L}$ relation of Becker et al. (\cite{BIT}) and
is 300K cooler than the Blue Edge. The resonance with the fourth overtone
occurs in sequences A and B at, respectively, $\mathrm{P}_1=4.02$\th day and
$\mathrm{P}_1=5.40$\th day. As already discussed, the calculations fail to
reproduce the jump of the {\it photometric} $\phi_{21}$. The variations
displayed by the models are never as large or as sharp as actually
observed. The situation is, however, entirely different in case of {\it
velocity curves}. For both sequences shown, the progression of velocity
$\phi_{21}$ has a shape remarkably similar to the observed one.  The
theoretical curves do not agree with the observations, though -- they are
displaced either to shorter or to longer periods in respect to the data.
It is easy to notice that the displacement depends on the position of the
$\omega_4 = 2\omega_1$ resonance within the sequence. Thus, a good match
between the models and the radial velocity data should be possible,
provided that the resonance period in the models is chosen properly.

\subsection{Position of the Resonance}

We now use the radial velocity $\phi_{21}$ data to constrain the position
of the s-Cepheid 2:1 resonance.  Fig.\th \ref{fig:phi21a} (bottom) already 
shows that it must be located somewhere between 4.02 and 5.40\th day (i.e.
between values for sequences A and B).  We want to pinpoint the resonance
center more precisely, though. To that aim we will try to "construct" the
model sequence that matches the velocity data as closely as possible. 

The hydrodynamical computations of Schaller \& Buchler (\cite{schaller})
show that the velocity $\phi_{21}$ for the overtone Cepheid models is very
tightly correlated with the resonant period ratio, 
$\mathrm{P}_4/\mathrm{P}_1$. In other words, to a very good approximation
we can write:

$$\phi_{21} = f(\mathrm{P}_4/\mathrm{P}_1)\eqno(3)$$

\ni where function $f$ is {\it the same for every model sequence}. Analogous 
property has also been found for the fundamen\-tal-mode Cepheids, where the 
$\omega_2\approx 2\omega_0$ resonance plays an important role (Buchler et~al.
\cite{BMK}). For {\it any} sequence of models, the relation between the 
period ratio and the period can be described by an approximate formulae

$$\mathrm{P}_4/\mathrm{P}_1=0.5\left(\frac{\mathrm{P}_1}{\mathrm{P}_\mathrm{r}}\right)^{-\alpha}\eqno(4)$$

\ni where $\mathrm{P}_\mathrm{r}$ is the period at the resonance center.
The parameters $\mathrm{P}_\mathrm{r}$ and $\alpha$ are different
for every sequence. Our goal is to find the values of these parameters,
for which a sequence reproduces the observations best. Substituting
Eq.\th (4) into Eq.\th (3) we get

$$\phi_{21}(\mathrm{P}_1) =
f\Biggl[0.5\left(\frac{\mathrm{P}_1}{\mathrm{P}_\mathrm{r}}\right)^{-\alpha}\Biggr]\eqno(5)$$

\ni $\mathrm{P}_\mathrm{r}$ and $\alpha$ can now be determined by fitting the
above expression to the observed values of velocity $\phi_{21}$.  This is done
with a $\chi^2$ method, minimizing

$$\chi^{2}(\mathrm{P}_\mathrm{r},\alpha)=\sum\left[\frac{\phi_{21}^{\mathrm{Obs}}(\mathrm{P}_1)-\phi_{21}(\mathrm{P}_1)}{\sigma(\phi_{21})}\right]^2\eqno(6)$$

\ni where $\phi_{21}^{\mathrm{Obs}}(\mathrm{P}_1)$ and 
$\phi_{21}(\mathrm{P}_1)$ are the observed and estimated values,
respectively, and $\sigma(\phi_{21})$ is the error of the observed
$\phi_{21}$.  For the function $f(\mathrm{P}_4/\mathrm{P}_1)$ we adopt the
$\phi_{21}$ progression of sequence A. It is the longest and most densely
sampled of all sequences of Schaller \& Buchler (\cite{schaller}).
\object{BD~Cas}, which deviates from the trend is omitted from the fit. 
Another star, \object{V440~Per}, turns out to be outside the range of
$\mathrm{P}_4/\mathrm{P}_1$ covered by the models. The minimization leads
to an excellent fit with $\chi^{2}_{\mathrm{min}}=18.97$, for $\chi^2$ of
20 degrees of freedom (22 stars, 2 parameters). The low value of
$\chi^{2}_{\mathrm{min}}$ shows that Eq.\th (5) indeed provides a good
representation of physical reality.  The values of the parameters
determined from the fitting procedure are $\alpha=0.141\pm 0.006$ and
$\mathrm{P}_{\mathrm{r}}=4.58 \pm 0.04$\th day ($1\sigma$ errors).

\begin{figure}
  \resizebox{\hsize}{13cm}{\includegraphics{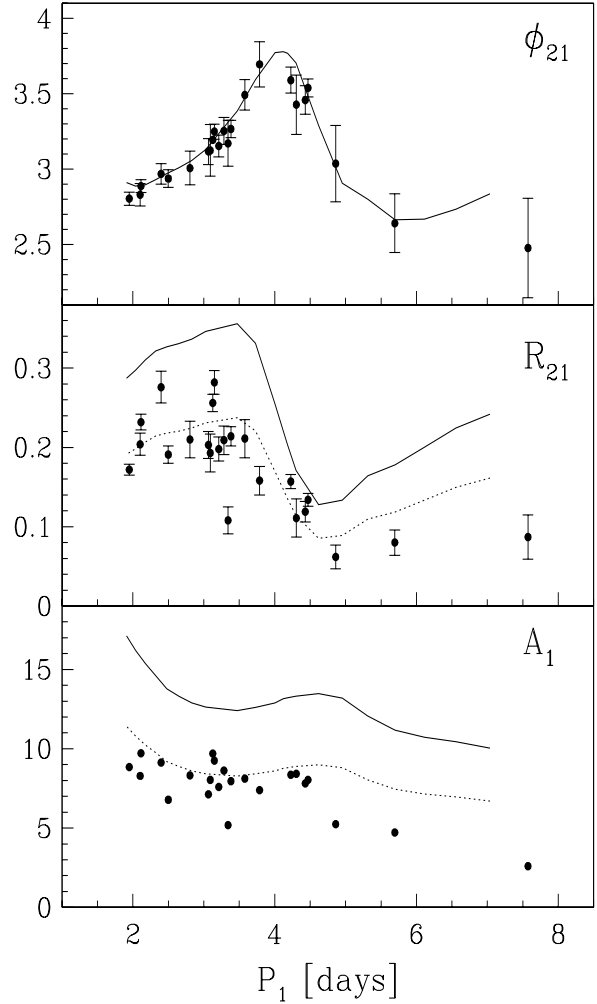}}
  \caption{Comparison of the s-Cepheid radial velocity observations with 
           the models of sequence A (see text): $\phi_{21}$ (top),
           $\mathrm{R}_{21}$ (middle) and $\mathrm{A}_1$ (bottom, in km/s). The 
           theoretical pulsation amplitude is scaled to the observed
           amplitude with the projection factor of $p=1.36$ (Burki et~al.
           \cite{burki}). The abscissa for the models is transformed 
           according to Eq.\th (4), with $\alpha=0.141$ and 
           $\mathrm{P}_\mathrm{r}=4.58$\th day. The dashed lines represent 
           $\mathrm{R}_{21}$ and $\mathrm{A}_1$ divided by constant 
           factor of 1.5 (see text). \object{BD~Cas} is omitted in all 
           plots.}
  \label{fig:fig2}
\end{figure}

The resultant $\phi_{21}$ fit is shown in Fig.\th \ref{fig:fig2} (top).
The plot confirms that a very good match between the models and the
$\phi_{21}$ data has been achieved. In the middle and bottom panels of the
figure we display the theoretical values of $\mathrm{R}_{21}$ and
$\mathrm{A}_1$ for the same sequence of models (solid lines).  Aside from
shifting and stretching of the abscissa (Eq.\th (4)) nothing else has been
adjusted here. The computed amplitude ratio $\mathrm{R}_{21}$ is
systematically too high as compared to the data (and so is $\mathrm{A}_1$),
but otherwise its progression with period bears a very strong resemblance
to the observations.  In particular, {\it both} the models and the data
display a steep decrease of $\mathrm{R}_{21}$ between periods of $3.5$ and
$5.0$\th day, followed by a slow increase.

The excessively high values of theoretical $\mathrm{A}_1$ and
$\mathrm{R}_{21}$ should not be consider a serious problem, in fact they
can be easily adjusted.  Both quantities depend sensitively on the
artificial viscosity, which is a numerical parameter in the hydrodynamical
codes.  By increasing the artificial viscosity, we can decrease
$\mathrm{R}_{21}$ of all models by a certain constant factor.  This will
also decrease the amplitude $\mathrm{A}_1$ by a similar (although somewhat
larger) factor. The velocity Fourier phase $\phi_{21}$, on the other hand, 
will remain almost {\it unchanged} and the good fit achieved for this
parameter will not be spoiled. This is a very general property of the
radiative hydrodynamical models (e.g.  Kov\'acs \cite{kovacs1}; Kov\'acs \&
Buchler \cite{kovacs2}; Kov\'acs \& Kanbur \cite{kovacs3}) and it holds for
the s-Cepheid models as well (Schaller \& Buchler \cite{schaller}).  The
dashed lines in Fig.\th \ref{fig:fig2} show theoretical $\mathrm{A}_1$ and
$\mathrm{R}_{21}$ of sequence A, both divided by the same factor of 1.5.
The scaled amplitude ratio $\mathrm{R}_{21}$ is in a very good {\it
quantitative} agreement with observations. Not only the location, but also
the size of the $\mathrm{R}_{21}$ drop is now reproduced.  The scaled
amplitude $\mathrm{A}_1$ is also much closer to the measured values, although
it is still somewhat too high.

Fig.\th \ref{fig:fig2} shows that the parameterization given by Eq.\th (4) can
bring velocity $\phi_{21}$ and $\mathrm{R}_{21}$ into remarkably good, {\it
simultaneous} agreement with the s-Cepheid observations. This proves that
the radiative models are capable of describing the radial velocity
variations in these stars.  It also shows that the 2:1 resonance plays a
dominant dynamical role in the s-Cepheid pulsations.  In that respect,
s-Cepheids are very similar to their fundamental mode counterparts. Indeed,
even the shapes of the velocity $\phi_{21}$ and $\mathrm{R}_{21}$
progressions are in both groups of stars remarkably alike (Buchler et~al.
\cite{BMK}; their Fig.\th 16).

The most important outcome of the fitting procedure is the determination of
the s-Cepheid resonant period. We stress, that this result is based {\it
solely} on matching the observed and computed $\phi_{21}$.  The derived
value is somewhat dependent on the way we define the function
$f(\mathrm{P}_4/\mathrm{P}_1)$. For example, if it is given by sequence B
instead of sequence A, we obtain $\mathrm{P}_{\mathrm{r}}=4.36\pm 0.03$\th
day. This is an extreme case, however. For other sequences of models we
{\it always} find higher values.  As our final estimate of the resonant
period we adopt the value resulting from fitting of sequence A.  We
consider it to be most trustworthy, because this particular sequence covers
the widest range of $\mathrm{P}_4/\mathrm{P}_1$ and simultaneously gives
the best match to the $\phi_{21}$ data (in terms of
$\chi^{2}_{\mathrm{min}}$).  We conclude, that the 2:1 resonance between
the first overtone and the fourth overtone occurs in s-Cepheids at a period
of $\mathrm{P}_{\mathrm{r}} = 4.58 \pm 0.04$\th day.

\subsection{Implications for the Mass-Luminosity Relation}

The evolutionary Mass-Luminosity relation for the classical Cepheids is
still a matter of considerable debate (cf. Buchler et~al. \cite{buchler}).
The newly derived position of the $\omega_4=2\omega_1$ resonance can be
used to constrain this important relation. In the following, we adopt the
slope of Becker et~al. (\cite{BIT}) and assume that the
$\mathrm{M}-\mathrm{L}$ relation has the form

$$\log(\mathrm{L}/\mathrm{L}_\odot) = 3.68\log(\mathrm{M}/\mathrm{M}_\odot) +
b\eqno(7)$$

\ni The zero point parameter $b$ has to be adjusted, so as to place the 
resonance at the right period. In order to determine the value of $b$, we
resort to the linear non-adiabatic (LNA) pulsation calculations.  We use the
linear pulsation code of Dziembowski (1977), with the
latest version of the OPAL opacities (Iglesias \& Rogers \cite{OPAL}) and
assume the standard Pop.\th I metallicity of $Z=0.02$.  The LNA
calculations show that the first overtone Blue Edge and the fundamental Blue
Edge are about 250~K apart. In the region between the two lines the models
can pulsate in the first overtone only. We assume, somewhat arbitrarily, that
the sample of Galactic s-Cepheids is represented best by the sequence of
models running on the H--R diagram parallel the two Blue Edges, half-way
between them.  For a sequence constructed in such a way, the resonance
condition ($\mathrm{P}_4/\mathrm{P}_1=0.5$ at $\mathrm{P}_1=4.58\pm
0.04$\th day) is satisfied for $b=0.73\pm 0.01$. 

The derived value of $b$ is not very far from the zero point of $b=0.65$,
inferred from the fundamental mode 2:1 resonance at $\mathrm{P}_0=10$\th
day (Moskalik \& Krzyt \cite{moskalik}, \cite{moskalik1}). The difference
of $\Delta b=0.08$ corresponds to 4.9\% difference of mass at a given
luminosity.  Although not in full agreement, the two values of $b$ are
close enough that with better models it might be possible to match both
resonances with the same $\mathrm{M}-\mathrm{L}$ relation. For the
s-Cepheid resonance located at 3.2\th day, as suggested in earlier
literature, a simultaneous match is hardly possible.  Such a low value of
$\mathrm{P}_\mathrm{r}$ would require a zero point of $b=1.05$, which is
incompatible with the fundamental mode constraint.

\section{Conclusions}

We have collected new CORAVEL radial velocity data and have then derived the
Fourier parameters of the pulsation velocity curves for 14 overtone
Cepheids. Our sample, combined with 10 variables of Krzyt et~al.
(\cite{krzyt}), covers the entire range of s-Cepheid periods. As such, it
is perfectly suited to discuss the group properties of this class of stars.
The main results of our work can be summarized as follows:

\begin{enumerate}
\item{}   {The progressions of the s-Cepheid Fourier parameters for velocity 
          curves and for light curves are very different. Velocity parameters 
          vary smoothly with the period and do not undergo any rapid changes 
          or jumps. The jump of the photometric $\phi_{21}$ at 
          $\mathrm{P}=3.2$\th day is not related to the resonance, but must 
          be caused by some other, as yet unidentified effect.}

\item{}   {Velocity Fourier parameters $\phi_{21}$ and $\mathrm{R}_{21}$ can 
          be reproduced remarkably well by the resonant (radiative) 
          hydrodynamical models. This clearly shows that the 2:1 resonance 
          with the fourth overtone is instrumental in shaping the s-Cepheid 
          pulsations. The fit of the models to the velocity data yields a
          new estimate of the resonance period, which is 
          $\mathrm{P}_\mathrm{r} = 4.58\pm 0.04$\th day. This period implies 
          the $\mathrm{M}-\mathrm{L}$ relation zero point of $b=0.73\pm 0.01$,
          not very different from value inferred from $\omega_2=2\omega_0$ 
          resonance.}

\item{}   Velocity Fourier parameters can discriminate between the overtone 
          and the fundamental-mode pulsators at all periods, including 
          $\mathrm{P} > 5.5$\th day. This property allows identification of 
          two new overtone Cepheids: \object{MY~Pup} ($\mathrm{P}=5.69$\th
          day) and V440~Per ($\mathrm{P}=7.57$\th day). The comparison with 
          the hydrodynamical models (cf. Fig.\th \ref{fig:fig2}) supports 
          this identification.
\end{enumerate}

The derived resonant period is based solely on the analysis of the
s-Cepheid radial velocity curves, in particular of the velocity Fourier
phase $\phi_{21}$, while the light curve information has been entirely
ignored.  There are two arguments in favor of such an approach.  First, the
velocity $\phi_{21}$ is predicted by the existing hydrocodes more robustly
than any Fourier parameter of the light curve.  It shows little
sensitivity to the the adopted artificial viscosity (as discussed above) or
to the choice of particular numerical scheme (e.g. Buchler, Kollath \&
Marom \cite{buchler4}).  In contrast to the light curve Fourier phases, it
also shows little sensitivity to the treatment of radiative transfer in the
optically thin outer layers of the star (Feuchtinger \& Dorfi
\cite{feuchtinger}).  This last property is particularly important, since 
most hydrocodes do a rather poor job in this respect. When velocity data
are available for comparison, the agreement between the models and
observations is always very good, much better than ever achieved for the
light curves.  This is clearly the case for the fundamental-mode Cepheids
(Moskalik et~al. \cite{MBM}). A very good match between computed
and observed overtone Cepheid velocity curves strengthens our confidence in the
above reasoning. The second argument comes from the fact that the resonant
interaction of pulsation modes is a {\it dynamical} phenomenon.  Since
$V_r$ is a {\it dynamical} quantity, studying its variations is, in our
opinion, the most direct and perhaps more basic way of probing resonance
effects in pulsating stars. 

We want to end this paper with a word of caution.  The hydrodynamical
models (Schaller \& Buchler \cite{schaller}; Antonello \& Aikawa
\cite{antonello3}) which can match the velocity data so successfully, at the 
same time fail to reproduce the observed light curves.  At this point we can
offer no explanation for this discrepancy. A new modeling effort with
better input physics is needed to address this problem and to confirm our
conclusions.  Such an effort, based on adaptive mesh hydrocode with 
time-dependent convection is already underway and its first results are
very promising (Buchler, private communication).

\begin{acknowledgements}
  We are grateful to G.~Schaller and J.-R.~Buchler for providing us with 
  their unpublished s-Cepheid models. This research has been supported in part
  by KBN (Poland) Grant No. 2-P30D-014-14. DB and FK gratefully acknowledge
  support from the Swiss National Fund for Scientific Research.
\end{acknowledgements}


\begin{thebibliography}{}

  \bibitem[1995]{alcock}
    Alcock, C., Allsman, R.A., Axelrod, T.S., et al., 1995, AJ, 109, 1653

  \bibitem[1993]{antonello2}
    Antonello, E., Aikawa, T., 1993, A\&A, 279, 119

  \bibitem[1995]{antonello3}
    Antonello, E., Aikawa, T., 1995, A\&A, 302, 105

  \bibitem[1986]{antonello1}
    Antonello, E., Poretti, E. 1986, A\&A, 169, 149

  \bibitem[1990]{antonello4}
    Antonello, E., Poretti, E., Reduzzi, L., 1990, A\&A, 236, 138

  \bibitem[1979]{baranne}
    Baranne, A., Mayor, M., Poncet, J.-L., 1979, Vistas Astron., 23, 279

  \bibitem[1995]{beaulieu}
    Beaulieu, J.-P., Grison, P., Tobin, W., et al., 1995, A\&A, 303, 137

  \bibitem[1977]{BIT}
    Becker, S.A., Iben, I., Tuggle, R.S., 1977, ApJ, 218, 633

  \bibitem[1994]{buchler3}
    Buchler, J.R., Buchler, N.E.G., 1994, A\&A, 285, 213

  \bibitem[1997]{buchler2}
    Buchler, J.R., Goupil, M.J., Piciullo, R., 1997, ApJ, 491, L99

  \bibitem[1996]{buchler}
    Buchler, J.R., Kollath, Z., Beaulieu, J.P., et al., 1996, 
    ApJ, 462, L83    

  \bibitem[1997]{buchler4}
    Buchler, J.R., Kollath, Z., Marom, A., 1997, Ap\&SS, 253, 139

  \bibitem[1986]{buchler1}
    Buchler, J.R., Kov\'acs, G., 1986, ApJ, 303, 749

  \bibitem[1990]{BMK}
    Buchler, J.R., Moskalik, P., Kov\'acs, G., 1990, ApJ, 351, 617

  \bibitem[1982]{burki}
    Burki, G., Mayor, M., Benz, W., 1982, A\&A, 109, 258

  \bibitem[1989]{chiosi}
    Chiosi, C., 1989, in IAU~Coll.~111 "The Use of Pulsating Stars in 
    Fundamental Problems of Astronomy", Ed.~E.G.~Schmidt (Cambridge\th :
    University Press), p.~19

  \bibitem[1977]{dziembowski}
    Dziembowski, W., 1977, Acta Astronomica, 27, 95

  \bibitem[1996]{feuchtinger}
    Feuchtinger, M.U., Dorfi, E.A., 1996, A\&A, 306, 837

  \bibitem[1990]{gieren}
    Gieren, W.P., Moffett, T.J., Barnes, T.G., et al., 1990, AJ, 99, 1196 

  \bibitem[1992]{gorynya1}
    Gorynya, N.A., Irsmambetova, T.R., Rostorgouev, et al., 1992,
    Pis'ma. Astron. Zh., 18, 777

  \bibitem[1996]{gorynya2}
    Gorynya, N.A., Samus, N.N., Rostorgouev, et al., 1996, 
    Pis'ma Astron. Zh., 22, 198

  \bibitem[1996]{OPAL}
    Iglesias, C.A., Rogers, F.J., 1996, ApJ, 464, 943.

  \bibitem[1981]{imbert}
    Imbert, M. 1981, IBVS 1983, 1

  \bibitem[1998]{kienzle}
    Kienzle, F., Pont, F., Bersier, et al., 1998, in "A Half 
    Century of Stellar Pulsation Interpretations: A Tribute to Arthur N. Cox",
    Eds.~P.A.~Bradley \& J.A.~Guzik, ASP~Conference Series, Vol.~135,~p.~241

  \bibitem[1990]{kovacs1}
    Kov\'acs, G., 1990, in "The Numerical Modelling of Nonlinear Stellar 
    Pulsations. Problems and Prospects", Ed.~J.R.~Buchler, NATO~ASI Series, 
    Vol.~302,~p.~73

  \bibitem[1989]{kovacs}
    Kov\'acs, G., Buchler, J.R., 1989, ApJ, 346, 898

  \bibitem[1993]{kovacs2}
    Kov\'acs, G., Buchler, J.R., 1993, ApJ, 404, 765

  \bibitem[1998]{kovacs3}
    Kov\'acs, G., Kanbur, S.M., 1998, MNRAS, 295, 834

  \bibitem[1996]{kovtyukh}
    Kovtyukh, V.V., Andrievsky, S.M., Usenko, et al., V.G., 1996,
    A\&A, 316, 155

  \bibitem[1999]{krzyt}
    Krzyt, T., Moskalik, P., Gorynya, et al., 1999, 
    (in preparation)

  \bibitem[1992]{mantegazza}
    Mantegazza, L., Poretti, E., 1992, A\&A, 261, 137

  \bibitem[1992]{metzger}
    Metzger, M.R., Caldwell, J.A.R., Schechter,P.L., 1992, AJ 103, 529

  \bibitem[1993]{moskalik2}
    Moskalik, P., Buchler, J.R., 1993, ApJ, 406, 190

  \bibitem[1992]{MBM}
    Moskalik, P., Buchler, J.R., Marom, A., 1992, ApJ, 385, 685

  \bibitem[1998]{moskalik}
    Moskalik, P., Krzyt, T., 1998, in "A Half Century of Stellar Pulsation 
    Interpretations: A Tribute to Arthur N. Cox", Eds.~P.A.~Bradley \& 
    J.A.~Guzik, ASP~Conference Series, Vol.~135,~p.~29

  \bibitem[1999]{moskalik1}
    Moskalik, P., Krzyt, T., 1999, (in preparation)

  \bibitem[1986]{petersen1}
    Petersen, J.O. 1986, A\&A, 170, 59

  \bibitem[1989]{petersen2}
    Petersen, J.O. 1989, A\&A, 226, 151

  \bibitem[1960]{petit}
    Petit, M., 1960, An Ap, 23, 681

  \bibitem[1994]{pont}
    Pont, F., Burki, G., Mayor, M., 1994, A\&AS, 105, 165

  \bibitem[1994]{poretti}
    Poretti, E., 1994, A\&A, 285, 524

  \bibitem[1994]{schaller}
    Schaller, G., Buchler, J.R., 1994, unpublished

  \bibitem[1991]{schmidt}
    Schmidt, E.G., 1991, AJ, 102, 1766

  \bibitem[1990]{simon4}
    Simon, N.R., 1990, MNRAS, 246, 70

  \bibitem[1986]{simon3}
    Simon, N.R., Aikawa, T., 1986, ApJ, 304, 249

  \bibitem[1981]{simon1}
    Simon, N.R., Lee, A.S., 1981, ApJ, 248, 291

  \bibitem[1985]{simon2}
    Simon, N.R., Moffett, T.J., 1985, PASP, 97, 1078

  \bibitem[1976]{simon}
    Simon, N.R., Schmidt, E.G., 1976, ApJ, 205, 162

  \bibitem[1995]{welch}
    Welch, D.L., Alcock, C., Bennett, D.P., et al., 1995,
    in IAU~Coll.~155 "Astrophysical Applications of Stellar Pulsation", 
    Eds.~R.S.~Stobie \& P.A.~Whitelock, ASP~Conference Series, Vol.~83,~p.~232
   
\end{thebibliography}
\end{document}